\documentclass[11pt,twoside,onecolumn]{article}
\usepackage[]{latexsym,amsmath,amssymb}
\usepackage[]{epsfig}
\newcommand{\be}{\begin{eqnarray}}
\newcommand{\ee}{\end{eqnarray}}

\title{\bf Some considerations on Determinism and Free Will \footnote{This essay has been completed at the University of Leiden, in October 2017}}
\author{Fabio~Scardigli$\,\,^{a,b}$\thanks{E-mail: fabio@phys.ntu.edu.tw}
\\
\\
{\em $^a$Dipartimento di Matematica, Politecnico di Milano,}\\
{\em Piazza Leonardo da Vinci 32, 20133 Milano, Italy.}\\
{\em $^b$Institute-Lorentz for Theoretical Physics, Leiden University,}\\ 
{\em P.O.~Box 9506, Leiden, The Netherlands.}}
\date{}
\begin{document}
\maketitle
\begin{abstract}
This article has been written, in a slightly different version, as an introductory chapter for the book collecting the essays of theoretical physicist Gerard 't Hooft, philosopher Emanuele Severino, and theologian Piero Coda, and inspired by the talks the three authors made as keynote speakers at the conference "Determinism and Free Will", held at the Cariplo Foundation Congress Center in Milan on May 13, 2017.

The conference was conceived and organized by a group of friends and colleagues consisting of Fabio Scardigli, Marcello Esposito, and Marco Dotti. We are grateful to our colleague Massimo Caccia, and especially to colleagues Gabriele Gionti and Massimo Blasone, for their help before and during the workshop.

The idea of organizing a meeting between Severino and 't Hooft had already been conceived several years ago. In fact, there had been a couple of unsuccessful attempts in 2010 and 2012 at the "DICE" Theoretical Physics conferences organized by Thomas Elze in Castiglioncello on a bi-annual basis. The opportunity arose in May 2017, when 't Hooft was in Castelgandolfo (Rome) for a conference organized at the Vatican Observatory and, on the way back to Holland, kindly agreed to stop for a couple of days in Milan, a city easily reachable by both Severino and Coda.

The conceptual reasons that led to this encounter lie first of all in the line of research pursued by 't Hooft for several years now, in which he aims to provide quantum mechanics with a deterministic foundation. His program seeks to bring this theory back under the umbrella of the most stringent determinism, a goal pursued by Einstein during the last decades of his life. On the other hand, Severino has built up an ontological vision that radically negates any reality in the \textit{becoming}, a point of view often associated with the strict deterministic conception of reality promoted by Einstein and Spinoza. He thus seemed to be the natural philosophical interlocutor for the physicist from Utrecht. Considering then the endless interweaving of the theme of free will with so many aspects of human experience, and also the happy accident of the 500th anniversary of the thesis presented by Luther (1517-2017), it seemed appropriate to complete the trio of speakers with the theologian Coda, who has always devoted a lot of attention to these issues.
%
\end{abstract}
\setcounter{equation}{0}

\newpage
%
\section{}
%
%
In Severino's vision, the \textit{becoming} (understood as the coming out of and the return to nothing of things) does not exist, i.e., it is not an element of reality. Becoming, far from being the most obvious, trivial, and undeniable evidence of the world, is indeed a theory, that is, just one 'interpretation' of events among the many possible. Indeed, Severino thinks that the interpretation of becoming, manifested since the Greek origins of Western thought as the oscillation of things between being and nothing, is just a "very stubborn illusion", a misinterpretation of events (words very similar to those used by Einstein to describe time, in a letter to the sister of his beloved friend Michele Besso, who had just passed away). With his philosophical research, Severino thinks he has provided a foundation for the eternity of beings, the eternity of each single entity, of each single event. This vision is undeniably similar to the vision proposed in general relativity, in which all events, past, present, and future, have always coexisted and will do so forever more, remaining eternally as points on the space-time manifold. The challenge for this point of view comes from the very heart of the other great theoretical construction of 20th century physics, quantum physics (at least perhaps until the recent studies of 't Hooft). Here in fact this vision clashes against Heisenberg's uncertainty principle, according to which the future is not strictly determined by the present, and the present is not strictly determined by the past, because there is a non-eliminable role played by chance in generating even elementary events. Physics, at least from the days of Maxwell and Boltzmann, has long been accustomed to using probabilistic laws to describe complex events, where it is reasonable to expect chance to play an important role. The novelty in the standard formulation of quantum mechanics was that even the elementary event, the absolutely simple event (think for example of a photon emitted by an electron in an atom) happens by pure chance. On the contrary, in the deterministic interpretation that 't Hooft proposes, quantum mechanics is instead brought back to the most complete, strict Einsteinian determinism. 
't Hooft's vision is thus somehow close to Severino's idea of the eternity of every single event, of the non-existence of becoming (which has always been thought of by Western philosophy as the random emergence of things from nothing).\\
%
%
%
\section{}
%
%
One of the main motivations of the 't Hooft program, viz., to render quantum mechanics (QM) strictly deterministic and therefore conceptually closer to general relativity (GR), lies precisely the fact that, once a greater conceptual homogeneity has been obtained between QM and GR (particularly as regards the idea of time advocated by the two theories), the much sought after goal of a unified theory of all physical phenomena would certainly be brought closer.

Indeed, if such a formulation exists, then QM would have a structure somehow similar to that of a classical theory, so it could be more easily re-formulated within the framework of general relativity. The unification of QM and GR would then, in principle, be far simpler and more natural.
 
The possibility, as shown by 't Hooft, of describing a system as simple as a cellular automaton, a perfectly classical and deterministic system, within the language of quantum mechanics, inevitably suggests that even the much more complicated system we observe, the physical world, so well described by that sophisticated quantum field theory called the Standard Model, may in fact be nothing but a very complicated, gigantic, deterministic cellular automaton.\\
%
%
%
%
\section{}
%
%
It has been said by several scholars that Severino's ontological vision appears to be an "influential metaphysics" of general relativity (to use Popper's locution), a sort of "general relativity" pushed to the extreme, with consistency and rigor. Severino seems in some respects stricter than Einstein when he establishes the eternity of every being. This vision naturally fills the "spaces" (social, psychological, economic, etc.) left necessarily empty by a physical theory, and the scenario is undoubtedly suggestive. However, Severino does not like to push too far the analogy between his position and the vision proposed by general relativity. In particular, he strongly emphasizes the different conceptual origins of the two logical structures. However, it must be said that the common features and the intrinsic coherence make it tempting to overlook the different origins of the two pictures. On the other hand, the scope and the terms used differ so much between them, that the existence of a channel of communication between the two structures appears to be almost miraculous.\\
%
%
%
%
\section{}
%
%
A possible critical point in the Severinian construction is his concept of "mathematical model" of the world. Severino says that, from its birth with Galileo and Newton, and until the end of the eighteenth century, modern science had an absolute, epistemic conception of the truth. Then, with the invention of ideas like non-Euclidean geometry and abstract algebra in mathematics, and with the quantum and relativistic revolutions in physics, the epistemic character disappeared, leaving room for the idea of science as hypothetical knowledge, designed to produce effective, working, and replaceable mathematical models of the world. This may suggest that the idea of "falsifiable" (mathematical) models appeared in physics and the other sciences only by the last two centuries. However, this view is not really corroborated by the most recent historical reconstructions. Lucio Russo and other scholars have shown with an abundance of detail that the concept of a mathematical model of a physical phenomenon was already developed with great clarity and effectiveness by the Alexandrians, and in particular Archimedes, Eratosthenes, Heron, and others.

So the idea that scientific theories are not absolute truths, i.e., not epistemic, but instead are (more or less efficient) mathematical models of the world has been in circulation for a much longer period than the last two centuries. Following the Alexandrians, the scientists of 16th and 17th centuries (starting with Galileo, but not forgetting the ideas of Leonardo da Vinci and others) subjected different physical theories (in the sense of models) to experimental tests in order to find out which one was the best, i.e., the most effective for representing the physical world. They didn't consider themselves to be hunting for absolute truth, but merely for a physical theory that was more effective than the old Aristotelian models. Hence, Newton set up his bucket experiment to test his idea of absolute space. And from the mid-17th century and during the 18th century, two different mathematical models of light, a wave theory and a corpuscular theory, were in fierce competition, with neither gaining the upper hand, until finally an experiment appeared to decide the issue (Young's double slit experiment).

It is therefore legitimate to state that falsifiable conceptions of scientific truth have always been present since the Alexandrian origins of modern science, or at least that they have long coexisted with the epistemic conceptions of scientific truth.\\
%
%
%
\section{}
%
%
%
The task of understanding the word "being" has long been believed to be a prerogative of pure theoretical philosophers. But today, in a world increasingly filled with the technological products of "quantum physics", it is almost unavoidable to ask "what lies behind this", perhaps with some kind of return to the origins, to the philosophy of nature put forward by Thales, Anaximander, Democritus, or Parmenides. The old quarrels between the founding fathers of quantum physics, which many physicists long considered a pastime for the "elderly" and irrelevant to the state of the art in physics, are once again being vigorously discussed. Physicists return to what they always should have been: philosophers of nature. Fundamental questions are being raised again, as when Einstein asked an astonished Abrahm Pais at the Princeton campus: "But are you really convinced that the Moon only exists if you look at it?" Of course, the underlying motivations are very often also prosaic and concrete. For instance, one conception rather than another of the quantum state, or of the problem of measurement, can completely change a whole line of research on quantum computers, and with it, the destination of millions of dollars of funding.\\
%
%
\section{}
%
%
With regard to QM, 't Hooft joins a long line of outstanding physicists who have shown discomfort with the standard interpretation of QM, or even criticized its foundations. It is well known that at least two of the founding fathers of QM, Einstein and Schr\"odinger, put forward critical insights into various aspects of the quantum point of view. And although they have generated research for about 80 years, many aspects of those problems remain without a shared consensus among the scientific community. Let us recall here just a few of these points:

a) For Einstein, QM is not a theory about single events. By definition, the fact that the theory has such a radically statistical structure prevents predictions about individual events (except for certain special cases): "The wave function $\psi$ does not describe, in any way, the condition of 'a' single system" (A.~Einstein, Physics and Reality, 1936).

b) In the famous EPR article (1935), Einstein claims to have demonstrated QM's "incompleteness": there are elements of physical reality that are not described, or captured, by the QM wave function.

c) Along the same lines, in the same year, Schr\"odinger launches another important idea in the form of his famous “cat paradox”. If we follow the standard interpretation of QM, in fact, before a direct observation ('measure') has been done, the cat in the box should be considered both alive and dead at the same time! Just as the radioactive atom (which controls the life of the feline through a clever mechanism) would result in a linear superposition of the decayed and non-decayed states.

d) For both Einstein and Schr\"odinger, the statistical character of QM, although captures a description of the reality with which each future model must be compared, it is not a good foundation upon which to build a theory able to describe single events, rather than just statistical descriptions of sets of events. Just as, according to Einstein, "the Newtonian laws of point particle mechanics cannot be inferred from thermodynamics" (Physics and Reality, 1936).

Einstein and Schr\"odinger's attitude towards QM is what the young Einstein, influenced by Mach, expressed with regard to the fundamental concepts of absolute space and time elaborated by Newton: "The prodigious success of his doctrine [Newtonian mechanics] obscured the critical investigation of its foundations [for two centuries]"(Herbert Spencer Lecture, Oxford 1933).\\
%
%
%
\section{}
%
%
%
An important topic of research in foundations of quantum mechanics directly involves the concept of free will, a concept which might seem, at first sight, to be linked to very concrete legal or social problems rather than to the foundations of an abstract physical theory.

In fact, one of the most debated (and paradoxical) results of quantum research in recent years is the so-called Free Will Theorem. This proceeds roughly as follows. First, the authors, Conway and Kochen, give a formal definition of free will which makes it possible to "quantify" the degree of "free will" possessed by a particular entity. Then, they analyze a Bell-type experiment (involving electron spin or photon spin/helicity), and demonstrate that, on the basis of commonly accepted QM principles, the observed electron (photon) must have the same degree of "free will" as the observer who performs the experiment.

The paradoxical and astonishing aspect of this conclusion is evident. How could an elementary particle (elementary, therefore without structure) have the same degree of free will as the human being who observes it? The real purpose of the theorem thus appears to be to highlight the paradoxical aspects of QM, rather like the Schr\"odinger cat experiment, but in another context.

For some, the content of the Free Will Theorem is even tautological. Indeed, if the world is completely deterministic, then neither the electron nor the observer have any free will because everything is completely predetermined. If, on the other hand, we admit that the observer has free will, then the world is not completely deterministic, and we seems to pay the price of seeing the electron exhibiting an indeterminacy, a "freedom" of choice, almost "its own free will".\\
%
%
%
%
\section{}
%
%
%
Bell's inequality is still the most frequently invoked argument against the possibility of building deterministic and local models of quantum phenomena. The vast majority of physicists believe that the lengthy debate triggered by Einstein's criticism in the 1930s has been definitively closed in favor of a non-deterministic interpretation of QM since the appearance of Bell's theorem in 1964. Those who propagate a return to determinism are often viewed as people (by now) far from the mainstream of scientific research. Nevertheless, some of the most original thinkers of our days, including 't Hooft, Penrose, Ghirardi, and others, have questioned various aspects of the standard Copenhagen interpretation of QM. Bell's inequality plays a key role in favor of the standard interpretation. However, the importance of the hypothesis of "measurement independence" in demonstrating the theorem was already clear to John Bell, and subsequently to other scientists like Shimony, Clauser, Horn, and others. This is an hypothesis that can be tied (and often is) to the "free will" of the observer who performs or oversees the measurement; that is to say, tied to the freedom of the observer to arbitrarily choose the orientation of the polarizing filters used in the measurement. On the decisive role played by this apparently innocent (and obvious) hypothesis, it is interesting to recall a memory by 't Hooft himself, according to which, during a meeting some thirty years ago, John Bell said: "If free will does not exist, then the deduction of the Bell inequalities is not valid."

In other words, the hypothesis of free will, or the observer's freedom of choice, is essential to the proof of Bell's inequalities. The latter are obeyed by any theory (with hidden variables) that is deterministic and local, and are violated by quantum mechanics. This is the standard argument that excludes a priori all local deterministic models of quantum phenomena involving hidden variables, since they do not violate Bell's inequalities, while QM does. Most people renounce deterministic local models in favor of quantum indeterminacy. However, Bell's inequality is clearly a consequence of the measurement independence hypothesis, which in turn can be naturally connected to the more than "obvious" assumption of freedom of choice for the observers themselves.

The use of the free will postulate (or equivalent assertions) to prove Bell's inequalities is confirmed also by the most recent formulations of such inequalities (see, for example, Brukner, Costa, Pikovski, Zych, "Bell Theorem for Temporal Order", arxiv:1708.00248). So, Bell's theorem and its (indirect) support for QM may appear as a kind of projection of the "obvious" hypothesis of attributing "free will" to human beings. Although it is not the only working model, QM appears instead under the weird light of being the model that fulfills our (natural) desire to attribute free will to us humans! One could almost say, in this subtle and specific sense, that QM is a "projection" of the human mind, owing to the dogma, which sounds typically Ptolemaic, of maintaining that humans possess the property of free will. These ideas fit well with those of the Free Will Theorem, whose authors claim, after giving a mathematical definition of the concept of free will, that if QM is true then the electron and its (human) observer have exactly the same degree of free will. A clearly absurd situation.\\
%
%
%
\section{}
%
%
%
The "hidden" but obvious hypothesis behind Bell's inequality is that of "measurement independence", closely related to the possibility of attributing freedom of choice (or free will) to the observer who performs or oversees the measurement. Somehow, since we humans want to have free will, we must therefore also attribute it to elementary particles. We cannot admit a deterministic description of the micro world, otherwise we too would be deterministic and we would not have free will. From this prospective QM looks almost like a “choice”. Humans want to have free will, so they naturally have to choose QM (which somehow guarantees it) over and above other models, which are discarded even though they could work (such as Bohmian mechanics, for example, at least in the non-relativistic regime), essentially because they are deterministic (and non-local).

Of course, the Severinian aspect of this situation will not have escaped the reader: we want, we believe, we choose to have free will. In a sense, we "choose" the world to be indeterminate to preserve our supposed free will; we somehow "choose" a world that is "becoming" (indeterminism) in order to better manipulate it. In this above-mentioned sense, the usual non-deterministic interpretation of QM looks rather like a "projection" of our mind.
The prevalence of a non-deterministic vision in the standard interpretation of quantum mechanics has been described by Severino in his book "Law and Chance" as a result of the more general course of Western philosophical thinking over the last two centuries. In Severino's words, "willpower 'wants' \textit{becoming} to exist, wants things to come out of nothing without a cause (randomly), to maximize the possibility of manipulating them". In some way, it wants standard QM to be the only proper representation of the physical world.\\
%
%
\section{}
%
%
%
The centrality of the hypothesis of freedom of choice for the observer is also emphasized by other authors. For example, Hossenfelder, in her blog, points out that if we deny the 
"free will" hypothesis, we lose Bell's theorem:
"The free will of the observer is a relevant ingredient in the interpretation of quantum mechanics. Without free will, Bell's theorem does not hold, and all we have learned from it goes out the window."\\
"The option of giving up free will in quantum mechanics goes under the name of 'superdeterminism' and is extremely unpopular. Unfortunately, it is highly probable that by scorning 'superdeterminism' we will miss something really important, something that could very well be a basis for future technologies."\\
"This kind of theories are often called 'conspiracy' theories, as it seems that the universe must be deliberately meant to prevent experimentalists from doing what they want. Therefore, this option is often not taken seriously."
"However, this could be a misleading interpretation of 'superdeterminism'. All that 'superdeterminism' means is that a state cannot be prepared independently of the detector settings. That doesn't necessarily imply a 'spooky' action at a distance, because the backwards light cones of the detector and state (in any reasonable universe) intersect anyway."

Of course, experiments have been planned (and some have already been done) that use very distant objects to close the loophole of "free will" or "measurement independence": stellar light experiments (Zeilinger, January 2017), and even with light emitted from quasars billions of light-years away (Kaiser, Gallicchio, planned for 2018). However it is clear that it is not possible to completely exclude an intersection between the particle light cone and that of the measuring instrument in the remote past. After all, it is assumed they both originate in the big bang.\\
%
%
%
\section{}
%
%
%
If we adopt the superdeterminist perspective, there is a problem in justifying the apparent, actual presence of "free will", of the ability to make "free" choices, that each of us experiences in everyday life. On this point it is interesting to report the ideas of Seth Lloyd, who shows that the illusion of "having" free will comes from the computational complexity of the processes (decision-making) that take place in our minds. A system, fundamentally deterministic but complex, is not able to predict its own decisions before taking them, because anticipating them has a degree of complexity similar to, or greater than, what is necessary to actually take them, and put them into practice. And this total blindness with regard to its own future choices, translates for the given system into the illusion of having freedom of choice, of exercising free will.
In fact, predicting ("calculating") what decision you will take in 10 minutes from now is such a complex and lengthy process, at least as complex and lengthy as actually taking it. From here arises the illusion of freely deciding. 

Free will reflects our ignorance of ourselves. Lloyd argued that free will is meaningful because our own decisions are unknown to us until we make them. Like all great theorems in computer science, his argument appeals to the paradoxes of recursion. When you think about yourself, you think about thinking about yourself, and you enter an endless loop: "Any system that can ask what it will be doing in five minutes' time, cannot always answer it. [...] It is less effective to simulate yourself, than it is simply to be yourself."

Lloyd summarizes this point of view on free will in a paper published in 2012, entitled 
\textit{A Turing test for free will}: "This article investigates the roles of quantum mechanics and computation in free will. Although quantum mechanics implies that events are intrinsically unpredictable, the pure 'stochasticity' of quantum mechanics adds only randomness to decision-making processes, not freedom. In contrast, the theory of computation implies that even when our decisions arise from a completely deterministic decision-making process, the outcomes of that process can be intrinsically unpredictable, even to - especially to - ourselves. I argue that this intrinsic computational unpredictability of the decision making process is what gives rise to our impression that we have free will."\\
%
%
%
\section{}
%
%
%
The interdisciplinary nature of the Milan conference meant that it was aimed at a wide audience, at bringing people together from many different backgrounds, so it was important not to limit the discussion only to the ontological status of determinism and free will. On the contrary, we tried also to investigate the connections that these concepts have with a variety of different human experiences, albeit in the limited space of one afternoon.

Of course, these interconnections would have required many different speakers to ensure that every point of view could be expressed. Clearly, it would have been a vain hope even just trying to achieve anything like completeness here, when we consider all the disciplines involved, i.e., biology, law, economy, neuroscience, history, theology, etc.

However, the rather relational nature of the theological point of view, as well as the happy contingency of the 500th anniversary of the publication of the Theses of Luther (1517-2017), suggested to us to complete the trio of speakers with the theologian Piero Coda.

In fact, the contribution of the Christian theological tradition to the formation of the very concept of freedom proper to modern Western civilization cannot be ignored. In his essay Coda illustrates how the concept of freedom emerges in the theological debate, in interaction with the concepts of grace and relation. This evolution is also vividly depicted through a comparison with the classical Greek concept of \textit{fate}, which envisaged, for humans and even for gods, a much more limited level of freedom in their actions, and therefore in their responsibilities, than what was instead predicated by the Christian doctrine in the first centuries of our era.

Of course, the languages and conceptual horizons of the three speakers were profoundly disparate, as the reader will easily realize by reading the book. However, these differences did not prevent an attempt at dialogue and interaction, an attempt that actually raises many more questions, rather than bringing many answers. But it seemed to us that this should in fact be one of the main objectives of an event conceived at the outset as multidisciplinary rather than specialized, and aimed at the general public.\\
%
%
%
\section{}
%
%
%
Some considerations that may perhaps also generate theological reflections are the following. A universe in which free will exists (not illusory, as described by Seth Lloyd, but fundamental) is necessarily a non-deterministic, a-causal universe in which at least some events (at least some in the mind of the person who makes the choice) happen, radically, by chance (otherwise free will, namely a choice free from the influences of past events, would not exist). On the contrary, a rigidly deterministic universe is guided by the most absolute and indisputable principle of causality. It is basically a universe whose entire history is given from the beginning, a universe in which "past and future would be equally present in the eyes of an intelligence equipped with superior analytic capabilities", as Laplace wrote. It would therefore be an 'eternal' universe, whose events are already all given and set from the beginning, and therefore in which time, in the sense of "becoming", is absent. This dualism has an interesting resemblance to the biblical myth of Adam and Eve's "fall" in the earthly paradise. Before choosing whether or not to eat the forbidden fruit, Adam and Eve lived in heavenly conditions, where all worries, pains, anguishes, and death were absent. A state of eternal idyllic present, a timeless condition of eternity. The choice of whether or not to eat the forbidden fruit, the exercise of free choice, destroys paradise and with it eternity. Human history, with all the joys and all the pains of the human condition, began at that moment, with that exercise of free choice. The biblical story, from this point of view, seems to link the beginning of human history to an act of free choice.\\
%
%
%
\section{}
%
%
%
Finally, it is worth reporting Vervoort's idea about the origin of probabilistic distributions as a product of underlying (micro) deterministic laws:\\
"It is straightforward to observe that any variable that is the effect of many hidden (and independent) causes, or in other words that is a function of many hidden and independent causes, is normally distributed, independently of the distribution of the causes (the latter may even have extremely simple, fully non-random-looking distributions). In other words, ‘(normal) probability’ is interpreted in terms of the occurrence of causes, more precisely, in terms of the massive averaging effect of a large number of causes. One is tempted to say that anything that has many causes will look probabilistic."

"If the majority of probabilistic properties of the world, namely those described by the normal distribution, can be understood as emerging from individual deterministic processes, it is tempting to conjecture that they all can.
There is a literally infinite number of probabilistic systems, from such diverse areas as fluid mechanics, diffusion, ballistics, error theory, population dynamics, population statistics, games of chance, quantum mechanics, information processing, and every field of engineering. All these profoundly different systems show the same frequency stabilization, the same need to converge towards well-defined ratios. They all obey the same simple laws of probability theory. The only possibility I can imagine to explain this 'necessity', shared by all these systems, is that they share the necessity of laws governing the evolution of their individual constituents, i.e., the necessity of determinism."\\
\section*{References}
\begin{enumerate}
%
\item
 G.~'t Hooft, \textit{The Cellular Automaton Interpretation of Quantum Mechanics},
   Springer Open, Berlin (2016).

\item
 E.~Severino, \textit{The Essence of Nihilism}, Verso Books, London (2016).

\item
 E.~Severino, \textit{Legge e Caso}, Adelphi Edizioni, Milano (1979).   

\item 
G.~'t Hooft, \textit{The Free Will Postulate in Quantum Mechanics}, 
   arXiv:quant-ph/0701097 (2007).

\item
L.~Russo, \textit{The Forgotten Revolution}, Springer Berlin (2004). 

\item
A.~Einstein, B.~Podolsky, N.~Rosen, \textit{Can quantum mechanical description of physical
   reality be considered complete?}, Phys. Rev. 47, 777 (1935).

\item
E.~Schroedinger, \textit{Die gegenwaertige Situation in der Quantenmechanik}, Die Naturwissenschaften 23, 807-812, 823-828, 844-849 (1935).

\item
A.~Einstein, \textit{Opere Scelte}, Bollati Boringhieri, Torino (1988).

\item
A.~Pais, \textit{Subtle is the Lord: The science  and the life of Albert Einstein}, Oxford University Press, Oxford (1982).

\item
J.~Conway, S.~Kochen, \textit{The Free Will Theorem}, Foundations of Physics,
   Vol.36, 1441 (2006).

\item
J.~S.~Bell, \textit{Speakable and Unspeakable in Quantum Mechanics}, Cambridge University
   Press, Cambridge (1987).

\item
M.~Zych, F.~Costa, I.~Pikovski, C.~Brukner, \textit{Bell’s theorem for temporal order},
   arXiv:1708.00248 (2017) 

\item
S.~Hossenfelder, \textit{Free will is dead, let’s bury it}, (2016)\\
    http://backreaction.blogspot.nl/2016/01/free-will-is-dead-lets-bury-it.html

\item
S.~Lloyd, \textit{A Turing test for free will}, Phil. Trans. Roy. Soc. A28, 3597 (2012). 

\item 
F.~Scardigli, \textit{A quantum-like description of the planetary systems}, Foundations of Physics, Vol.37, 1278 (2007).

\item
L.~Vervoort, \textit{Does Chance Hide Necessity?}, PhD Thesis, arXiv:1403.0145 (2014).
\end{enumerate}
%
%
\end{document}